\documentclass[aps,prl,twocolumn,showpacs,
tightenlines,superscriptaddress,floatfix]{revtex4}                                 \usepackage{graphicx}
\usepackage{dcolumn}
\usepackage{bm}
 
\begin{document}
 
\title{\bf On Raynal's posting in arXiv (nucl-th/0407060)}

\author{K. Amos}
\affiliation{School of Physics, University of Melbourne, Victoria 3010, Australia}
\author{L. Canton}
\affiliation{Istituto Nazionale di Fisica Nucleare, sezione di Padova,
e Dipartimento di Fisica dell'Universit\`a di Padova,
via Marzolo 8, Padova I-35131, Italia}
\author{G. Pisent} 
\affiliation{Istituto Nazionale di Fisica Nucleare, sezione di Padova,
e Dipartimento di Fisica dell'Universit\`a di Padova,
via Marzolo 8, Padova I-35131, Italia}
\author{J. P. Svenne}
\affiliation{Department of Physics and Astronomy, University of Manitoba,
and Winnipeg Institute for Theoretical Physics,
Winnipeg, Manitoba, Canada R3T 2N2}
\author{D. van der Knijff} 
\affiliation{Advanced Research Computing, Information Division,
University of Melbourne, Victoria 3010, Australia}

\date{\today}
\begin{abstract}
\sloppy

With nucl-th/0407060, Jacques Raynal uses the 
arXiv in a way which does not conform to standard professional practices.
His posting contains many statements that are beyond the borders of acceptable
scientific disputes, with the scope to defame colleagues by manifestly 
false or misleading statements.
In this comment we reject the three ``critiques'' expressed by Raynal.  
1. The fact that we possibly misquoted our references.
2. The role of the Pauli principle in these kind of calculations.
3. The nature and limits of our coupled-channel potential model.                   
Raynal's postings unfairly detract from the importance of our work,
which we published in Nuclear Physics A728, 65 (2003), on
a new approach, Multi-Channel-Algebraic-Scattering (MCAS), 
for coupled-channel calculations. With the MCAS approach we were
able to identify systematically all low-energy compound resonances,
and to include effectively the Pauli principle in collective, 
geometrical-type, macroscopic models of multichannel interaction.
This represents a clear advantage with respect to the current distribution 
of the ECIS formulation.
\end{abstract}


\maketitle              

Jacques Raynal, in his postings~\cite{1,2}, carries criticism of our work and of us
personally, and conducts slanderous and unprofessional behaviour, which 
raises some perplexities on the possible abusive usage of the popular 
arXiv service. Critiques, even strong criticisms, in science are very healthy
and, in general, can be accepted or rejected by confutation. They are part 
of the scientific method, and they contribute to the development of scientific 
knowledge. But critiques should adhere to the scientific 
issue, and provide new information that contributes to clarify the
subjects of dispute. Such type of critiques are respectful of the 
professionalism of the persons and refer only to the scientific debate.
A critique should not be used as a means to cast aspersion on colleagues
with false and misleading statements, in an attempt to manipulate or 
misrepresent factual matters for personal interests.

Below we deal with these problems in point
fashion. However, the reader should be aware that our disagreements were
founded upon a very simple item originally; to wit our use of a reduced
form of deformed spin-orbit interaction. Raynal's original objection
to our work was that we should have used the fuller derivation in
specifying channel interactions in a coupled channels analysis of
low-energy neutron $^{12}C$ scattering; his inference being that our overall
approach therefore was totally flawed. In the first posting~\cite{1} 
however, Raynal also initiated a personal attack and presented 
aspects of our work quite out of context. 
In our first arXiv response~\cite{3} we rejected by confutation his comments, 
providing our arguments about this very simple item of real 
debate of a scientific issue, but we also duly restated
the context and scope of our work, which was focused
on low-energy compound resonances and specifically on a systematic method of 
resonance computation and on the inclusion of Pauli principle in 
geometrical-collective-type approaches.

But the content of Raynal's new posting~\cite{2} goes well beyond that, 
and following discussion with the administrators of {\em arXiv.org}, 
we agreed that our present response to these tawdry affairs 
will be the final one. 

1. Raynal asserts that he cannot find in Hodgson's
book~\cite{4} justification of our claim~\cite{3} that Hodgson summarizes the
problem of using a deformed optical potential to reproduce a
low-energy spectrum of resonances. Raynal is wrong. Pages 426-427 of
Hodgson's book section (14.5) are devoted to the problem. It even has
the title Resonances with collective excitations. Therein Hodgson
describes and presents a very clear physical insight of the problems of
using a deformed optical model potential to reproduce the low-energy
spectrum of resonances. 

2. Raynal looked for discussion of the
anti-symmetrization problem in Hodgson's book, and could not find it for
scattering. He claims that in~\cite{3} we made such use of that reference, and 
concluded that we do not read the references that we quoted.
We never claimed that in Ref.~\cite{4} one can find a discussion on the 
anti-symmetrization problem for scattering.
In~\cite{3}, reference to Hodgson's book was solely related to the
summary of the problem of compound resonances and collective
excitations; the section 14.5 of Ref.~\cite{4}. 

3. Raynal writes that he cannot find what we claim to be in the other books 
that we quote in Ref.~\cite{3}. 
Incredibly Raynal also states that, as he could not
afford to buy the book by Greiner and Maruhn~\cite{5} 
(nor, apparently, access it in a library), he did not analyze it. 
How can he then dispute what we claim is contained therein? 
And in that book particularly the authors
note how coupling single particle dynamics with collective
(macroscopic-type) degrees of freedom of the target leads to violation
of the Pauli exclusion principle. Here we have to be very pedantic: 
that is precisely the statement to be found on page 297 of Ref.~\cite{5}, 
commencing on line 3.

4. Raynal states that he cannot find what we claim to be contained in the
book by Mahaux and Weidenmuller~\cite{6}. As above, the topic concerns
violation of the Pauli principle when coupling single particle dynamics
with collective-type degrees of freedom in the target. In that book on
page 103, one can find the sentence: ``In the papers concerned with 13 C as
a compound nucleus, the exclusion principle could not be exactly
satisfied". Similar sentences are to be found in pages~113-114.

5. On the basis of the assertions by Raynal of his inability to find that
which we attribute to be in the three books, he accuses us of not having
read from one or more of the books we quote. Such an accusation calls
into question our integrity and we vehemently reject that falsehood. 

6. The Mahaux and Weidenmuller book~\cite{6} deals with microscopic
calculations extensively, and the issue of anti-symmetrization is
seriously investigated. Nonetheless in that book, it is also stressed
that macroscopic model calculations, using collective
(geometrical-deformation) model excitations of the target, violate the
Pauli principle. So calculations with one-body type potentials, such as
are performed in the ECIS scheme, then have that flaw. The basic
calculation we have performed with the method described in Ref.~\cite{7} 
is exactly of that type as well. But in our case the effect of the Pauli
principle has been taken into account with the Orthogonalizing Pseudo
Potential (OPP) method. That method is applied in our approach by using
Eq.(43) in Ref.~\cite{7} and it works very well. That success, and the
demonstration in Ref.~\cite{7} that ignoring the Pauli effect gives much
spuriosity, belies Raynal's insistence that the Pauli principle, though
essential in dealing with bound many nucleon systems and (as he rightly
observes) is very important at higher energies, somehow has negligible
influence on low energy properties. 

7. Raynal attempts a stinging criticism of our work by drawing upon a 
comment made in Ref.~\cite{6}, namely that the study by Pisent (et al.) 
is an example of $^{12}C$ calculations with explicit violation of the 
exclusion principle. But that is exactly what we resolved in our recent 
paper~\cite{7} on which Pisent is one of the co-authors. 
The reference in the Mahaux and Weidenmuller book~\cite{6} is to
work of Pisent made in 1967. 

8. Raynal attempts to discredit our claim
of the importance of anti-symmetrization by drawing attention to a
paper, on which he is a co-author, as an example of 1 particle- 1 hole
$^{16}O(\gamma,n)$ calculation that ``treats anti-symmetrization correctly". 
Based on that work Raynal concludes that our claim of an anti-symmetrization
problem with the collective-model formulations, such as he has
programmed in his code ECIS, is not founded. 
That is illogical. The fact that Raynal and collaborators made a
``microscopic-type" calculation  (the definition of which we take as in the 
Greiner and Maruhn book, p.303), with the required overlap integrals 
to transform differential equations into integro-differential equations, 
in no way supports a conclusion that the ECIS formulation, for example, is free
from the methodological problems that we have raised. In fact the more
appropriate conclusion may well be the opposite with respect to what is
claimed by Raynal. To the best of our knowledge, the ECIS formulation is
not of a microscopic theory, nor in its current formulation is it
suitable for the solution of integro-differential coupled-channel
equations when allowance is made for adequate if not complete treatment
of the Pauli principle. As well, ECIS is not the basis of the
microscopic type calculations made in 1967 by Raynal (et al.). 
Even for the specific work of Raynal (et al.) so
referenced, it is noted in Ref.~\cite{6} on p.108, that their
(Raynal (et al.)) accord with experimental data at best is only
qualitative; something to be expected because ``the calculation contains
many approximations". With the above caveat and when the results shown
in Fig 6.1 of Ref.~\cite{6} are considered, it is clear that these 1967
results of Raynal (et al.) are inconclusive on the role of Pauli effects.

9. Certainly, Raynal is aware of the absolute need to treat
anti-symmetrization with nucleon scattering from nuclei, at least with
energies of 25 MeV or more. There is a great deal devoted to that in the
review of which he is a co-author~\cite{8}. In fact therein the
results found using Raynal's other large code, DWBA98, with and without
taking anti-symmetrization into account, and with a quite sophisticated
microscopic model of the reaction process, are very severe at all
energies to pion threshold and beyond. That he should now be dismissive
of the role of such at 0 to 5 MeV or more, is not supplemented by any
reasonable evidence.

10. In his criticisms~\cite{1,2}, Raynal also makes the
false claim that our model potential is already symmetric (before
symmetrization). Incidentally, had our potential been
already symmetric from the start, to symmetrize something already
symmetric is not a mistake. It would just be superfluous. But that was
not the case in the potential model we considered. 

11. Raynal acknowledges that the extra terms in the fuller description of the
deformed spin-orbit field plays a role primarily in the asymmetry of the
inelastic excitations. With that we concur and noted previously~\cite{3} that
the effects of the additional terms to the [L S] force have been studied
in the past and found to be of minor import; hardly to be seen to effect
with cross sections (inelastic ones) let alone to the elastic cross
sections for which it enters in second order. In fact a paper~\cite{9} on it
shows the effects of this extra term to be quite small in calculating
polarizations for all but forward scattering angles; and even there the
variations are not particularly large. 

12. Raynal contends that with a
refitting of parameters of our matrix of potentials, we could have fit
data without needing any OPP corrections. It is true that one may find
equivalent local potentials to any set of scattering phase shifts. In
the review on which Raynal is a coauthor~\cite{8}, there is a complete
chapter on methods and conditions for doing just that. But with
phenomenological potential forms the process is ''numerical inversion".
With such there is no serious constraint of physical nature as to what
interaction results. It is also true that by adjusting parameters
without constraint one can find interactions that separately fit elastic
and inelastic scattering data so one could equally well use Raynal's
argument to decry any phenomenological coupled-channel calculation. But
one cannot agree with Raynal's argument. A theoretical description should
produce sensible results on a variety of observables, and if the excited
bound and resonance spectra are so poorly reproduced in their character,
one must have first concern with the physics behind the specified model
system. Nevertheless, we did make serious attempts at parameter
variations in the first instance. We found that even with extreme values
for parameters we could never improve the situation with regard to
excessive spurious states. The fact is that ignoring the Pauli principle
always gives spurious results, while accounting for it leads quickly to
sensible results. 

13. The three conclusions reached by Raynal are false
because: 1) 
In our posting~\cite{3}, we have clearly specified what spin-orbit
potential we have used and what part of the fuller derivation that is.
That fact has never been hidden. 
2) We have given exactly what is written
and/or can be inferred from content in the three books under scrutiny.
We have specified in this letter where such can be found (giving page and
line numbers). In light of that, it is simply ridiculous to 
claim, as Raynal does so sarcastically and unprofessionally, that we did
not open one or another of those books. 
3) To state that, from Table 1 of
Ref.~\cite{7}, one cannot infer the importance of the Pauli principle is
erroneous. In that table, as is made quite clear in the text~\cite{7}, ``With
OPP" means that Pauli exclusion has been taken into account, ``Without
OPP" means that it has not and so is a result of a macroscopic form
factor calculation of the same ilk as might be made with ECIS, if such
could be used to generate sub-threshold and sharp resonance information.
It is clear and evident from the column labeled ``Without OPP" in the
table of relevance that there are many spurious states in both bound (E
$<$ 0) and scattering (E $>$ 0) regimes. In contrast, by treating the Pauli
effects (as we do) to get the results in the ``With OPP" column, the
number of sub-threshold bound and scattering states are closely aligned
to observed ones and they have the right spin-parities. It is important
to note that, without treating the Pauli principle, there are 8
additional spurious excited states and at least two spurious resonances.
It is also quite important that accommodating the Pauli principle does
not alter results, such as for the higher spin-parity states, for which
essentially there is no Pauli blocking since the relevant neutron orbit
is unoccupied in the target. That such is the case in our approach, with
the OPP corrections, is another validating feature. Instead this fact is
quoted erroneously by Raynal as evidence of irrelevance of the Pauli
effect. 

Finally the concentration by Raynal upon one small element in the
total scheme by which we solve the low-energy coupled channels problem
is disingenuous. At best it is a smoke screen. Raynal overlooks the
utility and adaptability of the method we have developed to analyze
low-energy scattering data whatever choice is made for the matrix of
interaction potentials. He also overlooks the fact that, despite our
choice for the spin-orbit field, that choice was made for convenience and
may be changed in future calculations without invalidating
the method used with a potentially more convenient form. Indeed with the MCAS
approach, a researcher may use whatever he/she opts for that matrix of
potentials. Nonetheless the specific choice we have made for the
low-energy model gives a matrix of potentials that is rich in structure.
The additional phenomenological spin-spin and orbit-orbit terms, taking
contributions to second order in the deformation, ensuring that the
matrix of potentials is hermitian, lead to results that are
overwhelmingly in agreement with mass 13 data. And, more importantly, we
have been able to account for the Pauli principle; something that is
imperative and not to be trivialized as Raynal has attempted in relation
with low-energy scattering. 
%

\end{document}